\documentclass[final,5p,times,twocolumn]{elsarticle}

\usepackage[colorlinks,allcolors={blue}]{hyperref}
\usepackage{amsmath}
\usepackage{natbib}
\usepackage{bm}
 
\usepackage{subfigure}
\usepackage{graphicx}
\usepackage{CJK}
\usepackage{xspace}  
\usepackage{multirow} 
\usepackage{dcolumn} 
\usepackage{pstricks}
\usepackage[utf8]{inputenc}
\usepackage{psfrag}
\usepackage[T1]{fontenc}
\usepackage{booktabs}
\usepackage{url}
\usepackage[normalem]{ulem}    

\newcommand{\ts}{\textsuperscript}
\newcommand{\tb}{\textsubscript}

\newcommand{\gammaray}{\ensuremath{\gamma} ray\xspace}
\newcommand{\gammarays}{\ensuremath{\gamma} rays\xspace}
\newcommand{\mevu}{MeV/nucleon\xspace}

\newcommand{\etwop}{\ensuremath{E(2^+_1)}\xspace}

\newcommand{\stwop}{\ensuremath{2^+_1}\xspace}

\newcommand{\beup}{\ensuremath{{B(E2)\!\!\uparrow}}\xspace}  

\newcommand{\esix}{1456(12)\xspace}
\newcommand{\eeight}{1115(34)\xspace}

\usepackage[normalem]{ulem}

\journal{Physics Letters B}

\begin{document}

\begin{frontmatter}



\title{Level Structures of \ts{56,58}Ca Cast Doubt on a doubly magic \ts{60}Ca}


\author[ahku,ariken,abeijing]{S.~Chen\corref{cor}} 
\cortext[cor]{Corresponding author, Present address: Department of Physics, University of York, York, UK} 
\ead{sdchen@hku.hk}
\author[ariken]{F.~Browne}
\author[ariken]{P.~Doornenbal}
\author[ahku]{J.~Lee}
\author[atudarmstadt,acea,ariken]{A.~Obertelli}
\author[acns]{Y.~Tsunoda}
\author[ariken,aut,ajaea]{T.~Otsuka}
\author[arcnp]{Y.~Chazono}
\author[aoakridge,atennessee]{G.~Hagen}
\author[atriumf,amcgill]{J.~D.~Holt}
\author[anccs,aoakridge]{G.~R.~Jansen}
\author[arcnp,aocu]{K.~Ogata}
\author[acns]{N.~Shimizu}
\author[ajaea,acns]{Y.~Utsuno}
\author[ajaea]{K.~Yoshida}
\author[acaen]{N.~L.~Achouri}
\author[ariken]{H.~Baba}
\author[acea]{D.~Calvet}
\author[acea]{F.~Ch\^ateau}
\author[ariken]{N.~Chiga}
\author[acea]{A.~Corsi}
\author[ariken]{M.L.~Cort\'es}
\author[acea]{A.~Delbart}
\author[acea]{J.-M.~Gheller}
\author[acea]{A.~Giganon}
\author[acea]{A.~Gillibert}
\author[acea]{C.~Hilaire}
\author[ariken]{T.~Isobe}
\author[atohoku]{T.~Kobayashi}
\author[ariken,acns]{Y.~Kubota}
\author[acea]{V.~Lapoux}
\author[acea,atudarmstadt,arit]{H.N.~Liu}
\author[ariken]{T.~Motobayashi}
\author[aijclab,ariken]{I.~Murray}
\author[ariken]{H.~Otsu}
\author[ariken]{V.~Panin}
\author[acea,abro]{N.~Paul} 
\author[ariken,ajaver,abogota]{W.~Rodriguez}
\author[ariken,aut]{H.~Sakurai}
\author[ariken]{M.~Sasano}
\author[ariken]{D.~Steppenbeck}
\author[acns,aatomki,aibs]{L.~Stuhl} 
\author[acea,atudarmstadt]{Y.L.~Sun} 
\author[arikkyo]{Y.~Togano}
\author[ariken]{T.~Uesaka}
\author[aut,ariken]{K.~Wimmer} 
\author[ariken]{K.~Yoneda}
\author[arit]{O.~Aktas}
\author[atudarmstadt,agsi]{T.~Aumann}
\author[ainst]{L.X.~Chung}
\author[aijclab]{F.~Flavigny}       
\author[aijclab]{S.~Franchoo}
\author[azagreb,atudarmstadt,ariken]{I.~Gasparic}
\author[auzk]{R.-B.~Gerst}
\author[acaen]{J.~Gibelin}
\author[aewha,aibs]{K.~I.~Hahn} 
\author[aewha,ariken,aibs]{D.~Kim} 
\author[aut]{T.~Koiwai}
\author[atoitec]{Y.~Kondo}
\author[atudarmstadt,agsi]{P.~Koseoglou}
\author[atudarmstadt]{C.~Lehr}
\author[ainst,avarn]{B.~D.~Linh}
\author[ahku]{T.~Lokotko}
\author[aijclab]{M.~MacCormick}
\author[auzk]{K.~Moschner}       
\author[atoitec]{T.~Nakamura}
\author[aewha,aibs]{S.~Y.~Park} 
\author[atudarmstadt]{D.~Rossi}
\author[aoslo]{E.~Sahin}
\author[atudarmstadt]{P.-A.~S\"oderstr\"om}
\author[aatomki]{D.~Sohler}
\author[atoitec]{S.~Takeuchi}
\author[atudarmstadt,agsi]{H.~T\"ornqvist}
\author[acsic]{V.~Vaquero}
\author[atudarmstadt]{V.~Wagner}
\author[aimpcas]{S.~Wang}
\author[atudarmstadt]{V.~Werner}
\author[ahku]{X.~Xu}
\author[atoitec]{H.~Yamada}
\author[aimpcas]{D.~Yan}
\author[ariken]{Z.~Yang}
\author[atoitec]{M.~Yasuda}
\author[atudarmstadt]{L.~Zanetti}

\address[ahku]{Department of Physics, The University of Hong Kong, Pokfulam, Hong Kong}
\address[ariken]{RIKEN Nishina Center, Wako, Saitama, Japan}
\address[abeijing]{School of Physics and State Key Laboratory of Nuclear Physics and Technology, Peking University, Beijing, China}
\address[atudarmstadt]{Institut f\"ur Kernphysik, Technische Universit\"at Darmstadt, Darmstadt, Germany}
\address[acea]{IRFU, CEA, Universit\'e Paris-Saclay, Gif-sur-Yvette, France}
\address[acns]{Center for Nuclear Study, University of Tokyo, RIKEN campus, Wako, Saitama, Japan}
\address[aut]{Department of Physics, University of Tokyo, Tokyo, Japan}
\address[ajaea]{Advanced Science Research Center, Japan Atomic Energy Agency, Tokai, Japan}
\address[arcnp]{Research Center for Nuclear Physics (RCNP), Osaka University, Ibaraki, Japan}
\address[atennessee]{Department of Physics and Astronomy, University of Tennessee, Tennessee, USA}
\address[atriumf]{TRIUMF, Vancouver, Canada}
\address[amcgill]{Department of Physics, McGill University, Montr\'eal, Canada}
\address[anccs]{National Center for Computational Sciences, Oak Ridge National Laboratory, Tennessee, USA}
\address[aoakridge]{Physics Division, Oak Ridge National Laboratory, Tennessee, USA}
\address[aocu]{Department of Physics, Osaka City University, Osaka, Japan}
\address[acaen]{LPC Caen, Normandie Univ., ENSICAEN, UNICAEN, CNRS/IN2P3, Caen, France}
\address[atohoku]{Department of Physics, Tohoku University, Sendai, Japan}
\address[arit]{Department of Physics, Royal Institute of Technology, Stockholm, Sweden}
\address[aijclab]{Universit\'e Paris-Saclay, CNRS/IN2P3, IJCLab, Orsay, France}
\address[abro]{Laboratoire Kastler Brossel, Sorbonne Universit\'e, CNRS, ENS, PSL Research University, Coll\`ege de France, Paris, France}
\address[ajaver]{Departamento de F\'isica, Pontificia Universidad Javeriana, Bogot\'a, Colombia}
\address[abogota]{Departamento de F\'isica, Universidad Nacional de Colombia, Bogot\'a, Colombia}
\address[aatomki]{Institute for Nuclear Research, Atomki, Debrecen, Hungary}
\address[aibs]{Center for Exotic Nuclear Studies, Institute for Basic Science, Daejeon, Korea}
\address[arikkyo]{Department of Physics, Rikkyo University, Tokyo, Japan}
\address[agsi]{GSI Helmholtzzentrum f\"ur Schwerionenforschung GmbH, Darmstadt, Germany}
\address[ainst]{Institute for Nuclear Science and Technology, VINATOM, Hanoi, Vietnam}
\address[azagreb]{Ru{\dj}er Bo\v{s}kovi\'c Institute, Bijeni\v{c}ka cesta 54, Zagreb, Croatia}
\address[auzk]{Institut f\"ur Kernphysik, Universit\"at zu K\"oln, K\"oln, Germany}
\address[aewha]{Department of Physics, Ewha Womans University, Seoul, Korea}
\address[atoitec]{Department of Physics, Tokyo Institute of Technology, Tokyo, Japan}
\address[avarn]{Vietnam Agency for Radiation and Nuclear Safety, Cau Giay, Hanoi, Vietnam}
\address[aoslo]{Department of Physics, University of Oslo, Oslo, Norway}
\address[acsic]{Instituto de Estructura de la Materia, CSIC, Madrid, Spain}
\address[aimpcas]{Institute of Modern Physics, Chinese Academy of Sciences, Lanzhou, China}


\begin{abstract}
Gamma decays were observed in \ts{56}Ca and \ts{58}Ca following quasi-free one-proton 
knockout reactions from \ts{57,59}Sc beams at $\approx200$\,\mevu. For \ts{56}Ca, a \gammaray transition was 
measured to be \esix\,keV, while for \ts{58}Ca an indication for a transition was observed at \eeight\,keV.
Both transitions were tentatively assigned as the $2^+_1 \rightarrow 0^+_{\mathrm{gs}}$ decays, and were compared to results 
from \textit{ab initio} and conventional shell-model approaches.
A shell-model calculation in a wide model space with a marginally modified effective nucleon-nucleon
interaction depicts excellent agreement with experiment for $2^+_1$ level energies, two-neutron
separation energies, and reaction cross sections, corroborating the formation of a new nuclear shell above the $N$\,=\,34 shell.
Its constituents, the $0f_{5/2}$ and $0g_{9/2}$\,orbitals, are almost degenerate.
This degeneracy precludes the possibility
for a doubly magic \ts{60}Ca and potentially drives the dripline of Ca isotopes to \ts{70}Ca or even beyond.
\end{abstract}




\end{frontmatter}


Understanding properties of atomic nuclei at the extremes, for example those with large proton-to-neutron imbalances, 
is of paramount importance in nuclear physics.
In these systems, often called exotic nuclei, new features emerge\,\cite{Nakamura:2017:PPNP}
including those that can be traced back to facets of nuclear forces. 
For instance, the tensor force, which has been known for decades\,\cite{Bethe:1940:PR,Bethe:1940:PR2},
can modify the spin-orbit energy splitting as a function of the proton number ($Z$) or the neutron number ($N$),
resulting in changes of shell structures, {\it i.e.}, shell evolution\,\cite{Otsuka:2005:PRL,Otsuka:2010:PRL}.
Examples have been found in several regions across the Segr\`{e} chart (see review papers,\,\cite{Sorlin:2013:PS,Otsuka:2020:RMP}).
Among them, the Ca isotopes provides an exemplary case of shell evolution,
with striking appearances of the new magic numbers $N$\,=\,32\,\cite{Huck:1985:PRC,Gade:2006:PRC,Wienholtz:2013:Nature} and
$N$\,=\,34\,\cite{Otsuka:2001:PRL,Steppenbeck:2013:Nature,Michimasa:2018:PRL,Chen:2019:PRL}.
The discovery of new magic numbers is usually followed by the exploration of the new nuclear shell lying above them, which may yield precious hints of the whereabouts of the dripline\,\cite{Erler:2012:Nature,Neufcourt:2019:PRL,Tsunoda:2020:Nature}.
This letter presents a finding along these lines based on state-of-the-art experimental and theoretical studies.

The Ca isotopes correspond to a complete filling of the $Z$\,=\,20 shell,
leading to a high sensitivity of the shell evolution according to the neutron number.
Signatures of magicity or sub-shell closures have been observed in the Ca isotopes 
at $N$\,=\,16,\,20,\,28,\,32,\,and\,34 based on the steep decrease of the two-neutron separation energies
$S_{2n}$\,\cite{Wienholtz:2013:Nature,Michimasa:2018:PRL} 
and the enhancement of the excitation energy of the first excited state
\etwop\,\cite{Huck:1985:PRC,Doornenbal:2007:PLB,Steppenbeck:2013:Nature,Raman:2001:ADNDT}.
The ground state of \ts{54}Ca has been shown, by knockout reactions,
to have a closed-shell configuration\,\cite{Chen:2019:PRL}, supporting the $N$\,=\,34 magicity.
Having the $N$\,=\,34 magic number thus confirmed, the nexus of interest is the shell above it.          
If the shell is composed only of the $0f_{5/2}$\,orbital, 
the recently observed \ts{60}Ca\,\cite{Tarasov:2018:PRL} may be doubly magic and become a dripline nucleus.
However, if the orbitals above $0f_{5/2}$ contribute substantially,
the dripline can be located deep into the {\it terra incognita} of the Segr\`{e} chart.
The influence of the $gds$\,orbitals above the $pf$\,shell is often discussed in the literature 
when neutron-rich Ca, Ti, and Ni isotopes are
addressed\,\cite{Poves:2001:NPA,Honma:2005:EPJA,Tsunoda:2014:PRC,Nowacki:2016:PRL,Taniuchi:2019:Nature,Cortes:2020:PLB}.
There are, however, no experimental data probing this shell
in the Ca isotopes.

Theoretical predictions of the level structure of Ca isotopes beyond \ts{54}Ca 
and the location of the dripline
have been made by modern shell-model, {\it ab initio}, beyond mean field 
calculations, 
and energy density functionals\,\cite{Poves:2001:NPA,Honma:2005:EPJA,Kaneko:2011:PRC,Holt:2012:JPG,Holt:2014:PRC,Hagen:2012:PRL,Coraggio:2014:PRC,Coraggio:2020:PRC,Li:2020:PRC,Magilligan:2021:PRC,Neufcourt:2019:PRL,Stroberg:2021:PRL}.
There seems to be no sign of convergence or consistency of such predictions for the level structure of \ts{56,58}Ca, as discussed later.
In fact, the predicted values of \etwop for \ts{56,58}Ca range from 0.5\,to
2\,MeV\,\cite{Kaneko:2011:PRC,Coraggio:2014:PRC,Coraggio:2020:PRC,Hagen:2012:PRL,Holt:2012:JPG,Li:2020:PRC,Magilligan:2021:PRC}.
Such a large variance prevents useful insights or conclusions regarding the shell structure
beyond the $N$\,=\,34 (sub-)shell closure. 
Recent predictions of a newly developed fitted interaction within the $fp$-model space, tailored for the neutron-rich Ca isotopes,
imply \ts{60}Ca being doubly magic at a similar level to \ts{68}Ni\,\cite{Magilligan:2021:PRC,Brown:2022:PHY}. 
This prediction is, however, strongly dependent 
on the agreement to experimental data for \ts{55--59}Ca\,\cite{Magilligan:2021:PRC,Brown:2022:PHY}. The
closest isotone along  $N$\,=\,40 with experimental information, \ts{62}Ti, showed no indication for 
a new magic number\,\cite{Cortes:2020:PLB}, in agreement with the predictions presented in Ref.\,\cite{Lenzi:2010:PRC}.
This letter reports on the first measurement of excitation energies of \ts{56,58}Ca
by means of in-beam $\gamma$-ray spectroscopy. Experimental data were 
confronted with modern shell-model and {\it ab initio} calculations combined with reaction theory.

The experiment was carried out at the Radioactive Isotope Beam Factory,
operated by the RIKEN Nishina Center and the Center for Nuclear Study, 
the University of Tokyo.
Radioactive beams were produced by fragmentation 
of a \ts{70}Zn beam at 345\,\mevu on a 10-mm-thick \ts{9}Be target. 
The \ts{57,59}Sc isotopes were then separated and identified from focal plane F0 to F13 of 
the BigRIPS separator\,\cite{Kubo:2012:PTEP}.
Afterwards, the secondary beams with intensities of 13.6\,particles/s for \ts{57}Sc
and 0.3\,particles/s for \ts{59}Sc impinged on the MINOS 
liquid-hydrogen (LH\tb{2}) target\,\cite{Obertelli:2014:EPJA} 
to induce proton knockout reactions. 
Reaction residues, \ts{56,58}Ca, were identified by the SAMURAI spectrometer\,\cite{Kobayashi:2013:NIMB}. 
Secondary beam energies at the target center were 
209\,\mevu for \ts{57}Sc and 199\,\mevu for \ts{59}Sc, inducing 
considerable Doppler shifts for the emitted \gammarays.
The DALI2\ts{+} detector array\,\cite{Takeuchi:2014:NIMA} was used
to measure the de-excitation \gammarays.
To overcome the large Doppler broadening partially caused by
the long LH\tb{2} target, Doppler corrections were performed
using the reaction vertex information reconstructed
by the MINOS time projection chamber.
For further experimental details, the interested reader is referred to
the supplemental material.

\begin{figure}[tb]
\centering
\includegraphics[scale=0.43]{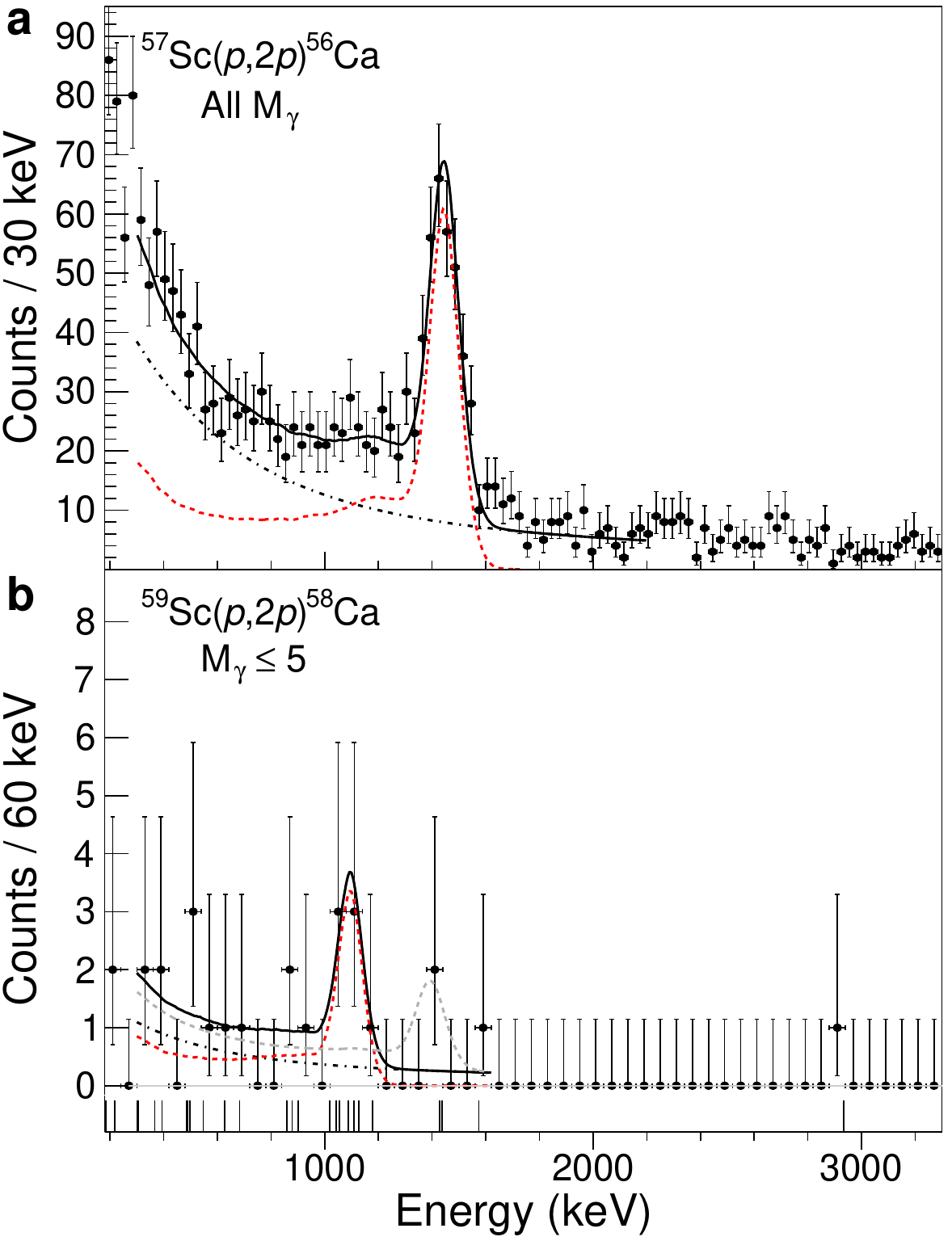}
\begin{picture}(0,0)
\put(-100,220){\includegraphics[scale=1.0]{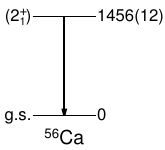}}
\put(-100, 90){\includegraphics[scale=1.0]{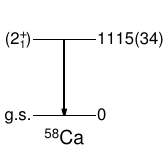}}
\end{picture}
\caption{\label{fig:dopp56Ca58Ca} Doppler-corrected $\gamma$-ray spectra. 
{\bf a}, Spectrum in coincidence with the \ts{57}Sc($p$,2$p$)\ts{56}Ca reaction. 
{\bf b}, Spectrum in coincidence with the \ts{59}Sc($p$,2$p$)\ts{58}Ca reaction restricted to $\gamma$\,multiplicity\,$\leq$\,5 
(spectra with other multiplicities are shown in Fig.\,3 of supplemental material).
Spectra were fitted with simulated DALI2\ts{+} response functions (red)
with a two-exponential background (black). 
Poisson-statistics errors were adopted for the data points. 
Unbinned data are shown in the bottom of panel {\bf b} for the \ts{59}Sc($p$,2$p$)\ts{58}Ca channel. 
The response curve for an assumed 1400\,keV $\gamma$-ray transition with a theoretical cross section of
0.25\,mbarn is indicated for \ts{58}Ca by the gray dashed line. }
\end{figure}

The Doppler-corrected $\gamma$-ray spectrum in coincidence with the \ts{57}Sc($p$,2$p$)\ts{56}Ca reaction 
is shown in Fig.\,\ref{fig:dopp56Ca58Ca}{\bf a}. 
A single peak is observed at 1456(12)\,keV and tentatively assigned to the
$2^+_1$\,$\rightarrow$\,$0^+_{\mathrm{gs}}$ transition. Energy uncertainties are dominated by the fitting error and energy calibration.
Lifetime effects on the measured energies were also evaluated. The heaviest
Ca isotope with a known \beup is \ts{50}Ca\,\cite{valiente:2009:PRL}. Assuming the same transition
strength, 37.5(10)\,$e^2$fm$^4$, for \ts{56}Ca gives a lifetime of 17\,ps. This lifetime value was
adopted with an error of 100\% and taken into account in the error determination.

Despite low statistics, the Doppler-corrected $\gamma$-ray energy spectrum
of the \ts{59}Sc($p$,2$p$)\ts{58}Ca, shown in Fig.\,\ref{fig:dopp56Ca58Ca}{\bf b},
revealed a peak-like structure in the energy range of 1000--1200\,keV.
To test the significance level of this peak, a maximum likelihood fit procedure was applied to 
the unbinned data (see bottom of Fig.\,\ref{fig:dopp56Ca58Ca}{\bf b}). 
The background in this spectrum was modeled from the \ts{57}Sc($p$,2$p$)\ts{56}Ca reaction,
with the amplitude normalized according to the event numbers.
This procedure was validated with the \ts{55}Sc($p$,2$p$)\ts{54}Ca data from the same experiment,
which yielded a good description of the background.
A significance of 2.8\,$\sigma$, defined as the peak amplitude over the statistical
uncertainty from the maximum likelihood fit,
was obtained for the tentative \eeight\,keV $\gamma$-ray
transition, including systematic errors from lifetime effects, and tentatively assigned 
to the $2^+_1$\,$\rightarrow$\,$0^+_{\mathrm{gs}}$ decay of \ts{58}Ca. 
The assumed lifetime is 66\,ps based on the same assumption as \ts{56}Ca.
Noteworthy are the two counts observed at $\sim$1400\,keV,
comparable to the \etwop of \ts{56}Ca. However, taking into account calculated theoretical
cross sections, as discussed below, resulted in a poor overall agreement of the response function with 
the data, as evidenced by the gray dashed line in Fig.\,\ref{fig:dopp56Ca58Ca}{\bf b}.
Further tests for the validity and the impact of the \ts{58}Ca data is discussed in the supplemental material.

\begin{table*}[tb]
    \centering
    \caption{\label{tab:cx}Observed excitation energies ($E$\tb{exp}) in keV and cross sections ($\sigma$\tb{exp}) in mbarn from the
    \ts{57}Sc($p$,2$p$)\ts{56}Ca and \ts{59}Sc($p$,2$p$)\ts{58}Ca reaction channels compared to theoretical
        values ($\sigma$\tb{th}) using the DWIA calculated single-particle cross sections ($\sigma$\tb{sp})
        and spectroscopic factors ($C^2S$) from VS-IMSRG, GXPF1Bs, and A3DA-t. Predicted spin-parities ($J^\pi$),
        associated proton-removal orbitals ($nl_j$), and excitation energies ($E_x$) are also provided. 
        }
    \begin{tabular}{ccccccccccccccc}
        \midrule
        \midrule
        &\multicolumn{2}{c}{Experiment} & & & DWIA & \multicolumn{3}{c}{VS-IMSRG} & \multicolumn{3}{c}{GXPF1Bs} & \multicolumn{3}{c}{A3DA-t} \\
        \cline{2-3} \cline{7-9} \cline{10-12} \cline{13-15}
        \vspace{-0.3cm}\\                                
        &$E$\tb{exp} & $\sigma$\tb{exp} & $J^\pi$ & $nl_j$ & $\sigma_\mathrm{sp}$ & $E_x$ & $C^2S$\tb{th} & $\sigma$\tb{th} & $E_x$ & $C^2S$\tb{th} & $\sigma$\tb{th}  & $E_x$ & $C^2S$\tb{th} & $\sigma$\tb{th} \\
        \midrule
         \multirow{3}{*}{\ts{56}Ca} & 0         & 0.80(6) & $0^+_\mathrm{g.s.}$ & $0f_{7/2}$ & 1.80 & 0    & 0.61 & 1.10 & 0    & 0.69 & 1.24 & 0    &  0.62 & 1.11 \\
                                    & \esix     & 0.43(4) & $2^+_1$             & $0f_{7/2}$ & 1.74 & 1002 & 0.29 & 0.50 & 1416 & 0.25 & 0.44 & 1519 &  0.27 & 0.47 \\
                                    &           &         & $4^+_1$             & $0f_{7/2}$ & 1.73 & 1307 & 0.05 & 0.09 & 1776 & 0.02 & 0.04 & 2339 &  0.01 & 0.02 \\
                                    & Inclusive & 1.23(5) &                     &           &      &      &      & 1.69 &      &      & 1.72 &      &       & 1.60 \\
        \midrule
         \multirow{3}{*}{\ts{58}Ca} & 0         & 0.66(24)& $0^+_\mathrm{g.s.}$ & $0f_{7/2}$ & 1.58 & 0    & 0.80 & 1.26 & 0    & 0.83 & 1.31 & 0     & 0.46 & 0.73 \\
                                    & \eeight   & 0.47(19)& $2^+_1$             & $0f_{7/2}$ & 1.54 & 1075 & 0.16 & 0.25 & 1382 & 0.15 & 0.23 & 1040  & 0.42 & 0.65 \\
                                    &           &         & $4^+_1$             & $0f_{7/2}$ & 1.52 & 1423 & 0.001& 0.002& 1772 & 0.001& 0.002& 2084  & 0.05 & 0.08 \\
                                    & Inclusive & 1.14(15)&                     &           &      &      &      & 1.51 &      &      & 1.54 &       &      & 1.46 \\
        \midrule
        \midrule
    \end{tabular}
\end{table*}

%

\begin{figure}[!tb]
\centering
\includegraphics[scale=1.0]{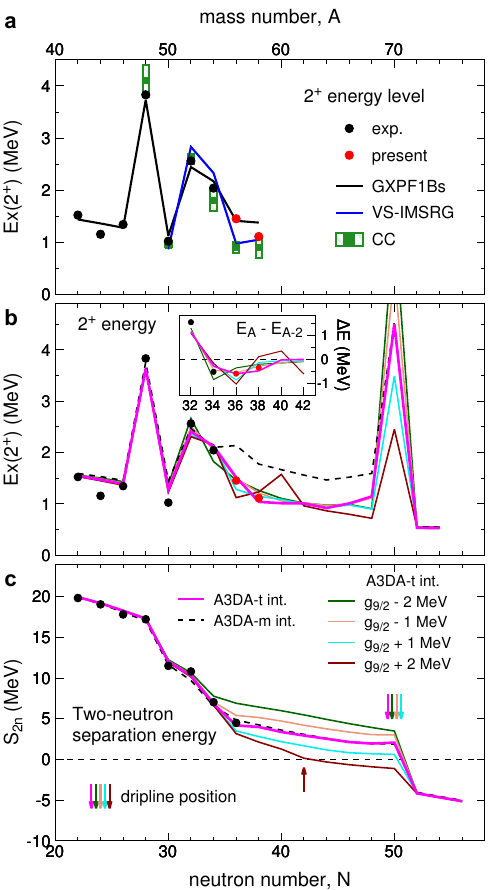}
\caption{\label{fig:CaE2S2n} Comparison of calculated \etwop and $S_{2n}$ values with experimental data.
{\bf a}, \etwop systematics in even-even Ca isotopes
confronted with theoretical approaches: The shell model using the GXPF1Bs 
Hamiltonian, the VS-IMSRG method, and CC calculations. 
{\bf b}, \etwop systematics in even-even Ca isotopes, and their differences (inset). Experimental points are the same as {\bf a}.
The calculated values are obtained by the original A3DA-m Hamiltonian as well as its revised one (A3DA-t)
{\bf c}, $S_{2n}$ systematics in even-even Ca isotopes. Also shown in {\bf b} and {\bf c} are the effect of shifting the neutron
$0g_{9/2}$\,orbital for predictions of A3DA-t.
}
\end{figure}

The systematics of \etwop values as a function of neutron number presented in Fig.\,\ref{fig:CaE2S2n}{\bf a} evince 
the expected pattern for magic nuclei at $N$\,=\,28: A sharp increase from $N$\,=\,26 to 28 
followed by a large reduction at $N$\,=\,30. Similarly, a characteristic sharp increase from 
$N$\,=\,30 to 32 exists for the $N$\,=\,32 magic number, while the enhanced \etwop at $N$\,=\,34 
is indicative of magicity. 
The firmly established data point for \ts{56}Ca and the tentative one for \ts{58}Ca 
are as low as the $N$\,=\,22,\,24,\,26,\,and\,30 values, with a decrease from $N$\,=\,36 to 38.

The \etwop systematics of Ca isotopes were compared to conventional shell-model calculations with the GXPF1Bs
Hamiltonian in the model space of the full $pf$ shell\,\cite{Chen:2019:PRL,Honma:2005:EPJA}, and two state-of-the-art {\it ab initio} approaches: 
The valence-space in-medium similarity renormalization group (VS-IMSRG)
\,\cite{Tsukiyama:2012:PRC,Hergert:2016:PR,Stroberg:2017:PRL,Stroberg:2019:ARNPS}
and the coupled-cluster theory (CC)\,\cite{Hagen:2014:RPP},
both employing the two- (N\!N) and three-nucleon (3N) interaction 1.8/2.0\,(EM)\,\cite{Hebeler:2011:PRC},
derived from chiral effective field theory\,\cite{Machleidt:2011:PR}.
Details of these theoretical approaches are provided in the supplemental material.
Figure\,\ref{fig:CaE2S2n}{\bf a} shows the theoretical calculations well describe the \etwop 
excitation energies up to $N$\,=\,34, and the GXPF1Bs Hamiltonian also provides
a good agreement with the present experimental value for \ts{56}Ca.
However, all these calculations predict a flat behavior from $N$\,=\,36 to 38.

A more general discussion provides an instructive viewpoint 
of the \etwop values of \ts{56,58}Ca.
If the $0f_{5/2}$\,orbital is isolated from the other orbitals, 
$N$\,=\,36 corresponds to a system of two neutrons solely occupying the $0f_{5/2}$\,orbital.
Likewise, $N$\,=\,38 would be four neutrons in the $0f_{5/2}$\,orbital, 
or, equivalently, two neutron holes of the fully occupied $0f_{5/2}$\,orbital.
The two-body interaction is invariant between particle and hole systems,
but the single-particle energies can vary with neutron number. 
Such changes of single-particle energies do not affect excitation level energies, because only one orbital is relevant.
Thus, assuming the $0f_{5/2}$ neutron orbital is marginally modified between \ts{56}Ca and \ts{58}Ca, the \etwop value 
should be identical between $N$\,=\,36 and $N$\,=\,38 as a consequence of this particle-hole symmetry.
It is emphasized that this consequence is independent of the choice of the two-body interaction.

The present results indicate a decrease of \etwop from $N$\,=\,36 to 38 by several hundred keV.
This observation conflicts with the arguments above, implying a non-isolated $0f_{5/2}$\,orbital.
Since all experimental evidence supports an $N$\,=\,34 magic number in the Ca isotopes, the $0f_{5/2}$\,orbital is considered 
as isolated from lower-energy orbitals. This points to the other possibility that the $0f_{5/2}$\,orbital is coupled to higher orbitals,
suggesting a shell comprising the $0f_{5/2}$\,orbital and at least one higher orbital.
This new shell has never been discussed and
its appearance excludes the $N$\,=\,40 magic number in Ca isotopes.

A previous 
theoretical study has discussed an ``$sdg$'' shell
built on an inert \ts{60}Ca core\,\cite{Nowacki:2016:PRL}. 
This approach proved valid for the \ts{78}Ni region, but remains untested for the Ca isotopes. In the present work,
only the characteristics of the $0g_{9/2}$ and $1d_{5/2}$\,orbitals can be constrained by experiment. 
As the protons can be assumed to form a $Z$\,=\,20 closed shell,
only neutrons above $N$\,=\,20 are treated as valence nucleons in the calculations.

The existing effective A3DA-m\,\cite{Tsunoda:2014:PRC} $NN$ interaction, defined for a model space 
comprising the full $pf$ shell, the $0g_{9/2}$, and $1d_{5/2}$\,orbitals, is used as a starting point.
It has been successfully used for the systematic descriptions
of Ni ($Z$\,=\,28)\,\cite{Tsunoda:2014:PRC} and Cu ($Z$\,=\,29)\,\cite{Ichikawa:2019:naturephy} isotopes.
Figure\,\ref{fig:CaE2S2n}{\bf b} shows that the observed \etwop values are well reproduced
by the A3DA-m interaction up to $N$\,=\,34, and substantial deviations
for $N$\,=\,36 and 38. The excitation-energy lowering from $N$\,=\,36 to 38 is
well reproduced by the A3DA-m interaction, in contrast to trends shown in Fig.\,\ref{fig:CaE2S2n}{\bf a}.
This suggests a minor revision of the interaction may be sufficient to reproduce the experimental data.
Figure\,\ref{fig:CaE2S2n}{\bf c} shows the $S_{2n}$ values for the Ca isotopes. 
There is no notable deviation for the nuclei where experimental data are available,
implying the validity of the A3DA-m interaction.

The A3DA-m interaction is revised by varying only the two-body-matrix-elements (TBME)
in the linear combination (LC) method\,\cite{Chung:1976:Phdths,Honma:2002:PRC}
to better reproduce the \etwop values of \ts{54,56,58}Ca.
Changes to the TBMEs are small, as expected. 
The maximum change is 0.198\,MeV, while the others are much smaller.
The correlation between the original TBMEs and the revised TBMEs are shown in the supplemental material.

The revised interaction is labeled ``A3DA-t'' hereafter.
Figure\,\ref{fig:CaE2S2n}{\bf b} depicts the \etwop values obtained with the A3DA-m and the A3DA-t interactions
from \ts{42}Ca to \ts{74}Ca, with A3DA-t reproducing the \etwop values of \ts{56,58}Ca.
The \etwop value remains almost constant until \ts{68}Ca, where the value for \ts{70}Ca rises 
due to the filled $0f_{5/2}$-plus-$0g_{9/2}$ shell and the necessity of neutron excitations to 
the high-lying $1d_{5/2}$\,orbital.
As orbitals above this are not included, the present work cannot describe excitation energies much 
beyond \ts{70}Ca.

Figure\,\ref{fig:CaE2S2n}{\bf c} shows $S_{2n}$ values up to \ts{76}Ca,
the last possible nucleus in the present model space. A plateau is formed from \ts{56}Ca to
\ts{70}Ca. Beyond this, $S_{2n}$ becomes negative, implying the dripline is located at \ts{70}Ca,
close to some predictions\,\cite{Koura:2005:PTP,Erler:2012:Nature}, 
beyond others\,\cite{Hergert:2020:FP} or within argued ranges\,\cite{Tarasov:2018:PRL,Neufcourt:2019:PRL,Stroberg:2021:PRL}.
Inclusion of higher $sdg$\,orbitals may slant the dripline even further
due to quadrupole correlations\,\cite{Nowacki:2016:PRL,Taniuchi:2019:Nature}.
As noted in Ref.\,\cite{Nowacki:2016:PRL}, neutrons in higher $sdg$\,orbitals can enhance quadrupole collectivity, 
which may lead to a well-deformed ground state of \ts{70}Ca.

The sensitivity of the neutron $0g_{9/2}$\,single-particle-energy (SPE)
was characterised by varying it up to $\pm$2\,MeV with respect to the original value
of the A3DA-t interaction. Results of this can be seen in Fig.\,\ref{fig:CaE2S2n}{\bf b}
for the \etwop and in Fig.\,\ref{fig:CaE2S2n}{\bf c} for the $S_{2n}$. Of particular interest
is the difference of \etwop, defined as $\Delta{E}$\,=\,$E(2^+_A)$\,-\,$E(2^+_{A-2})$ and shown in the inset of Fig.\,\ref{fig:CaE2S2n}{\bf b}. The larger
the neutron $0g_{9/2}$\,SPE, the larger the drop from \ts{54}Ca to \ts{56}Ca, producing a local \etwop maximum
for \ts{60}Ca and shifting the dripline to \ts{62}Ca. A positive shift of +1 or +2\,MeV can be excluded from the
experimental \etwop and $S_{2n}$ of \ts{56}Ca. Conversely, a too low $0g_{9/2}$\,SPE
value quenches the experimentally established magicity at $N$\,=\,34\,\cite{Steppenbeck:2013:Nature,Michimasa:2018:PRL,Chen:2019:PRL},
resulting in a high neutron $0g_{9/2}$ occupation number not observed in \ts{54}Ca\,\cite{Chen:2019:PRL}. 
Our results obtained from \ts{56}Ca challenge the notion of an $N$\,=\,40 magicity
at \ts{60}Ca and are reinforced by the tentative experimental value for \ts{58}Ca, as all $\Delta{E}$ 
remain negative except for the $0g_{9/2}$\,SPE shifted by +2\,MeV.

\begin{figure}[!tb]
\centering
\includegraphics[scale=0.7]{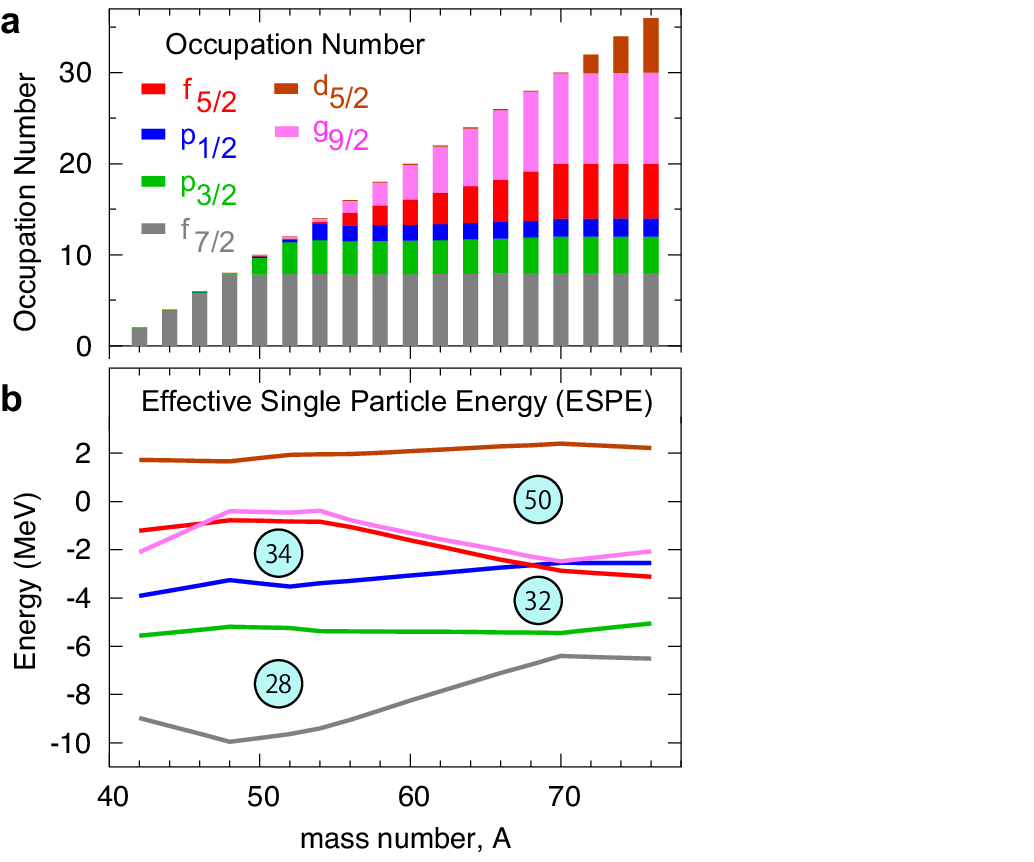}
\caption{\label{fig:occ_espe} 
{\bf a}, Occupation numbers of the ground states of even-even Ca isotopes obtained from the A3DA-t interaction. 
{\bf b}, Effective single-particle energies for the same states as panel {\bf a}.
The numbers in the circles are the neutron numbers corresponding to the 
magic gaps.
}
\end{figure}

The occupation number of each single-particle orbital is shown in Fig.\,\ref{fig:occ_espe}{\bf a}.
Likewise, the ESPE\,\cite{Otsuka:2020:RMP} are displayed in Fig.\,\ref{fig:occ_espe}{\bf b}.
The shell structure above $N$\,=\,34 is clearly characterized by two orbitals,
$0f_{5/2}$ and $0g_{9/2}$, which remain almost degenerate across the range shown in Fig.\,\ref{fig:occ_espe}{\bf b}.
Thus, the emergence of a new shell above $N$\,=\,34 comprising $0f_{5/2}$, $0g_{9/2}$\,orbitals
and some others, like $1d_{5/2}$, is evident.
The $N$\,=\,34 magic gap decreases beyond $A$\,=\,60, but has little affect 
because the $1p_{1/2}$\,orbital remains almost completely occupied.
This degeneracy is lifted for isotopes with $Z$\,>\,20 due to the strong monopole attraction
between a proton in $0f_{7/2}$ and a neutron in $0f_{5/2}$\,\cite{Otsuka:2020:RMP}.

Further discussions are concentrated on the cross sections.
Inclusive cross sections for 
the \ts{57}Sc($p$,2$p$)\ts{56}Ca and \ts{59}Sc($p$,2$p$)\ts{58}Ca reactions were 
measured to be 1.23(5) and 1.14(15)\,mb, respectively. Partial cross sections of 
the excited states were extracted using the efficiency-corrected $\gamma$-ray intensities,
and those to the ground-states deduced by subtraction. All measured cross 
sections are summarized in Tab.\,\ref{tab:cx}. 
Inclusive and partial cross sections were comparable between both nuclei, 
lending support to the assignment of a peak in \ts{58}Ca.

Theoretical cross sections were obtained by combining 
single-particle cross sections $\sigma$\tb{sp} calculated from the distorted-wave impulse approximation (DWIA)
and the spectroscopic factors $C^2S$ from the GXPF1Bs and A3DA-t Hamiltonians, and VS-IMSRG approach described above\,\cite{Duguet:2015:PRC}.
They are listed in Tab.\,\ref{tab:cx}. 
The beams of ground-state \ts{57,59}Sc have assumed $J^\pi$\,=\,$7/2^-$.
Only removal from the proton $0f_{7/2}$\,orbital was considered, as higher-lying proton orbitals contributed
only a few percent to the final states. Negligible cross sections were 
calculated to states other than the listed $0^+_\mathrm{g.s.}$, 
$2^+_1$, and $4^+_1$ states. 

Similar inclusive cross sections for both reaction channels were
predicted, as observed experimentally, and with $\sigma$\tb{exp}-to-$\sigma$\tb{th} ratios $\sim$0.75, 
agreeing with previous values obtained in the 
region\,\cite{Cortes:2020:PLB,Sun:2020:PLB,Liu:2019:PRL} and for stable nuclei\,\cite{Aumann:2021:PPNP}. This signifies a low 
occupation number of protons across the $Z$\,=\,20 shell in the ground states of 
\ts{57,59}Sc, hence a good proton shell closure. A different picture
is observed for the partial cross sections to the \stwop states. While
the $\sigma$\tb{exp}-to-$\sigma$\tb{th} ratio holds for \ts{56}Ca, despite considerable  
uncertainties, the experimental partial cross section for \ts{58}Ca
is two times larger than the value predicted by the GXPF1Bs and VS-IMSRG calculations.
In contrast, the A3DA-t Hamiltonian gives results in good agreement
with experiment: Partial cross sections change from $N$\,=\,36 to 38 in a consistent manner with experiment.

In conclusion, the first spectroscopy measurements for
\ts{56,58}Ca following the 1$p$-knockout reactions from scandium isotopes were carried out.
A \gammaray transition associated with the $2^+_1$\,$\rightarrow$\,$0^+_{\mathrm{gs}}$
decay was assigned for \ts{56}Ca and an indication of this transition was observed for \ts{58}Ca.
A comparison with standard shell-model and {\it ab initio} theoretical calculations
exhibits a notable deficiency in their descriptions of nuclear structure around $N$\,=\,40. 
The particle-hole symmetry argument robustly leads to a new shell comprising at least the $0f_{5/2}$ and $0g_{9/2}$\,orbitals 
above $N$\,=\,34. The fitted A3DA-t interaction, introduced in this work, 
shows an excellent description of so far known experimental data,
and predicts the Ca dripline at $N$\,=\,50, because of substantial correlation energies from the pairing
between the $0f_{5/2}$ and $0g_{9/2}$\,orbitals. 
Thus, the picture of the new magic number $N$\,=\,34\,\cite{Otsuka:2001:PRL,Steppenbeck:2013:Nature,Michimasa:2018:PRL,Chen:2019:PRL} 
becomes more complete with a shell built atop of it.
More detailed structure information of \ts{56,58}Ca can be obtained 
by future measurements with Ge detector arrays at next-generation facilities like FRIB\,\cite{FRIB},
notably validation of the tentative transition at \eeight\,keV.  
Additional experimental investigations of neutron-rich Ca isotopes,
such as particle states in \ts{55}Ca from neutron pickup reactions
and the spectroscopy of \ts{60}Ca, not believed to be doubly magic from the assessment of results presented here
and Ref.\,\cite{Cortes:2020:PLB}, would provide key information to
further characterize the new shell.






\appendix

We would like to express our gratitude to the RIKEN Nishina Center accelerator staff for providing the stable and
high-intensity primary beam and to the BigRIPS team for operating the secondary beams.
S.C. acknowledges the support of the IPA program at RIKEN Nishina Center.
F.B. acknowledges the support of the Special Postdoctoral Researcher Program.
J.L. acknowledges the support from Research Grants Council (RGC) of Hong Kong with grant of Early Career Scheme (ECS-27303915).
K.O. acknowledges the support from JSPS KAKENHI Grants No. JP16K05352.
Y.U. and N.S. acknowledge the support from JSPS KAKENHI Grant No. 20K03981.
Y.L.S. acknowledges the support of the Marie Sk{\l}odowska-Curie Individual Fellowship (H2020- MSCA-IF-2015-705023).
V.V. acknowledges support from the Spanish Ministerio de Econom\'ia y Competitividad under Contract No. FPA2017-84756-C4-2-P.
L.X.C. and B.D.L. would like to thank MOST for its support through the Physics Development Program Grant No. {\DJ}T{\DJ}LCN.25/18.
H.N.L. acknowledges the support from the Deutsche Forschungsgemeinschaft (DFG, German Research Foundation) - Project No. 279384907- SFB 1245. 
I.M. acknowledges the hospitality of RIKEN through the IPA program.
R.-B.G. was supported  by the Deutsche Forschungsgemeinschaft (DFG) under Grant No. BL 1513/1-1.
D.R. and V.~Werner acknowledge the Deutsche Forschungsgemeinschaft (DFG, German Research Foundation) under grant SFB 1245.
V.~Werner and P.K. acknowledge the German BMBF grant number05P19RDF N1.
P.K. was also supported by HGS-HIRe.
D.~Sohler was supported by projects No. GINOP-2.3.3-15-2016-00034 and No. NKFIH-K128947.
I.G. was supported by HIC for FAIR and Croatian Science Foundation under projects no. 1257 and 7194.
K.I.H., D.K., and S.Y.P. acknowledge the support from the IBS grant funded by the Korea government (No. IBS-R031-D1). 
This work was also supported 
by the United Kingdom Science and Technology Facilities Council (STFC) under Grants No. ST/P005314/1 and No. ST/L005816/1,
by JSPS KAKENHI Grant Nos. JP16H02179 and JP18H05404, and by NKFIH (128072).
The development of MINOS was supported by the European Research Council through the ERC Grant No. MINOS-258567.
J.D.H. acknowledges support from TRIUMF, which receives funding via a contribution through the National Research Council of Canada, NSERC, and S.~R.~Stroberg for the imsrg++ code\,\cite{imsrgplusplus} used to perform the VS-IMSRG calculations.
G.H. and G.R.J acknowledges support by the US Department of Energy under desc0018223 (NUCLEI SciDAC-4 collaboration) and under Contract No. DE-AC05-00OR22725 with UT-Battelle, LLC (Oak Ridge National Laboratory). Computer time was provided by the Innovative and Novel Computational Impact on Theory and Experiment (INCITE) program. This research used resources of the Oak Ridge Leadership Computing Facility and of the Compute and Data Environment for Science (CADES) located at Oak Ridge National Laboratory, which is supported by the Office of Science of the Department of Energy under Contract No. DE-AC05-00OR22725. 
The A3DA calculations were performed on the supercomputer Fugaku at RIKEN AICS and Oakforest-PACS operated by JCAHPC (hp190160, hp200130, hp210165).
This work was supported in part by MEXT as ``Priority Issue on Post-K computer'' (Elucidation of the Fundamental Laws and Evolution of the Universe) and ``Program for Promoting Researches on the Supercomputer Fugaku'' (Simulation for basic science: from fundamental laws of particles to creation of nuclei) and by JICFuS.
This work was supported in part by MEXT KAKENHI Grant No. JP19H05145.

\bibliography{local.bib}

\end{document}